\documentclass[superscriptaddress,10pt,onecolumn,showpacs,amsmath,amssymb]{revtex4}
\usepackage{amsfonts}
\usepackage{amssymb}
\usepackage{amsmath}
\usepackage{epsfig}
\usepackage{graphicx}
\usepackage{dcolumn}
\usepackage{bm}
\usepackage{float}
\usepackage{graphicx}

\usepackage{color}
\newcommand{\beq}{\begin{equation}}
\newcommand{\eeq}{\end{equation}}
\newcommand{\beqs}{\begin{eqnarray}}
\newcommand{\eeqs}{\end{eqnarray}}

\newcommand{\be}{\begin{equation}}
\newcommand{\ee}{\end{equation}}
\newcommand{\bea}{\begin{eqnarray}}
\newcommand{\eea}{\end{eqnarray}}

\begin{document}

\title{The effects of massive graviton on the equilibrium between the black hole and radiation gas in an isolated box}

\author{Ya-Peng Hu}\email{huyp@nuaa.edu.cn}
\address{College of Science, Nanjing University of Aeronautics and Astronautics, Nanjing 210016, China}
\address{Instituut-Lorentz for Theoretical Physics, Leiden University, Niels Bohrweg 2, Leiden 2333 CA, The Netherlands}
\address{Key Laboratory of Theoretical Physics, Institute of Theoretical Physics, Chinese Academy of Sciences, Beijing 100190, China}

\author{Feng Pan}\email{fan_physics@126.com}
\address{College of Science, Nanjing University of Aeronautics and Astronautics, Nanjing 210016, China}

\author{Xin-Meng Wu}\email{wuxm@nuaa.edu.cn}
\address{College of Science, Nanjing University of Aeronautics and Astronautics, Nanjing 210016, China}
\begin{abstract}
It is well known that the black hole can has temperature and radiate the particles with black body spectrum, i.e. Hawking radiation. Therefore, if the black hole is surrounded by an isolated box, there is a thermal equilibrium between the black hole and radiation gas. A simple case considering the thermal equilibrium between the Schwarzschild black hole and radiation gas in an isolated box has been well investigated previously in detail, i.e. taking the conservation of energy and principle of maximal entropy for the isolated system into account. In this paper, following the above spirit, the effects of massive graviton on the thermal equilibrium will be investigated. For the gravity with massive graviton, we will use the de Rham-Gabadadze-Tolley (dRGT) massive gravity which has been proven to be ghost free. Because the graviton mass depends on two parameters in the dRGT massive gravity, here we just investigate two simple cases related to the two parameters, respectively. Our results show that in the first case the massive graviton can suppress or increase the condensation of black hole in the radiation gas although the $T-E$ diagram is similar like the Schwarzschild black hole case. For the second case, a new $T-E$ diagram has been obtained. Moreover, an interesting and important prediction is that the condensation of black hole just increases from the zero radius of horizon in this case, which is very different from the Schwarzschild black hole case.

\end{abstract}



\maketitle

%\newpage

\section{Introduction}
Among the research on the black hole physics, a significant progress is the discovery that the black hole can has temperature and radiate the particles with black body spectrum, i.e. Hawking radiation~\cite{Hawking:1974rv}. Note that, before the discovery of Hawking radiation, there have already been several clues shown that the black hole may exists the thermodynamics like a thermal system. For example, the area $A$ of the horizon of a general black hole could never decrease found by Hawking \cite{Hawking:1971tu}, which is even argued to relate to the entropy of black hole by Bekenstein~\cite{Bekenstein:1973ur}. Moreover, four mechanical laws can be found for the the stationary black hole, which are very similar like the thermodynamical laws of thermal system~\cite{Bardeen:1973gs}. After the discovery, one will find that the four mechanical laws are just the four laws of thermodynamics for a stationary black hole system, and hence the thermodynamics of a stationary black hole is constructed.

It should be pointed out that the Hawking radiation is a quantum effect, since black hole usually absorbs particles classically. Therefore, one simple consequent question is that whether there is some equilibrium between the black hole and radiated particles if the black hole is surrounded by an isolated box, i.e. an equilibrium between the absorbed and radiated particles~\cite{Gibbons:1976pt}. For the radiated particles, usually they can be considered as the radiation gas whose energy and entropy are well known. Therefore, the above question can be also equivalent to whether there is some stable condensation of a black hole among the radiation gas in an isolated box. For the simple case, the condensation of a black hole is just the Schwarzschild black hole, which has been well investigated in detail in the pioneer work by G.W Gibbons and M.J Perry~\cite{Gibbons:1976pt}. Their most impressive results are that indeed there is some equilibrium between the Schwarzschild black hole and radiation gas in an isolated box. Moreover, the equilibrium condition has been analytically obtained after taking the conservation of energy and principle of maximal entropy for the isolated system into account~\cite{Gibbons:1976pt}.

In this paper, we would like to investigate the effect of massive graviton on this equilibrium. A simple motivation is that the massive gravity has been paid many attentions recently~\cite{Hinterbichler:2011tt,deRham:2014zqa}. It has been found that the massive gravity can be not only a theoretical proposal, but also a possibility to interpret the dark matter and dark energy problem~\cite{Hinterbichler:2011tt,deRham:2014zqa}. In addition, according to the AdS/CFT correspondence, the effect of massive graviton in the bulk gravity can be related to the effects from the lattice in the dual field theory, i.e. deducing the momentum dissipation of electrons~\cite{Vegh:2013sk,Blake:2013owa,Davison:2013jba,Amoretti:2014zha,Hu:2015dnl}. It should be emphasized that the extension from Einstein's gravity to a massive gravity with massive graviton is difficult. The reason is that many massive gravity theories suffer from the instability problem of the Boulware-Deser ghost~\cite{Hinterbichler:2011tt,deRham:2014zqa,Fierz:1939ix,Boulware:1973my}. Recently, the so called de Rham-Gabadadze-Tolley (dRGT) massive gravity has been proposed~\cite{deRham:2010ik,deRham:2010kj}, which is a nonlinear massive gravity theory and has been found to be ghost free~\cite{Hassan:2011hr,Hu:2015xva,Zhang:2015nwy}. Therefore, we will use the dRGT massive gravity to investigate the effect of massive graviton on the equilibrium between the black hole and thermal radiation gas in an isolated box. For the simplicity, we also just take the static Schwarzschild-like black hole in the dRGT massive gravity into account~\cite{Cai:2014znn,Hendi:2015pda,Hendi:2015hoa,Cao:2015cza,Hu:2016hpm}. After taking the conservation of energy and principle of maximal entropy for the isolated system into account, we investigate the effects of massive graviton on the equilibrium. Note that, since the graviton mass just depends on two parameters in the four dimensional spacetime case, and hence in our paper we only consider two simple cases which are related to the two parameters, respectively. Furthermore, these two simple cases can be analytically investigated. The main interesting results are that the $T-E$ diagram is similar like the Schwarzschild case in the first case, while a new $T-E$ phase diagram has been obtained in the second case. Moreover, in the first case, we can further find that the massive graviton can suppress or increase the condensation of black hole in the radiation gas, which depends on the value of the other parameter $c_2$.

The rest of our paper is organized as follows. In Sec.~II, we give a brief introduction to the pioneer work by G.W Gibbons and M.J Perry as a warmup to make the whole paper more readable and completeness. In Sec.~III, we use the dRGT massive gravity to investigate the effects from massive graviton on the equilibrium, and consider two simple cases which can be analytically investigated in detail. Sec. IV is devoted to the conclusion and discussion.

\section{Warmup: Equilibrium between the Schwarzschild black hole and radiation gas in an isolated box}
Suppose that there is an isolated box of volume $V$ contained with the radiation gas, and the total energy for this isolated system is $E$ (More details of this box can be seen in Ref.~\cite{Gibbons:1976pt}). For the radiation gas, the energy and entropy are well known as
\begin{equation}
E_g=aVT^4,~~S_g=\frac{4}{3}aVT^3, \label{EERS}
\end{equation}
where $a$ is the Stefan's constant, and the subscript $g$ is related to the radiation gas. For the Schwarzschild black hole,
\begin{equation}
ds^2=-(1-\frac{2M}{r})dt^2+(1-\frac{2M}{r})^{-1}dr^2+r^2(d\theta^2+\text{sin}^2\theta d \varphi^2), \label{SchBMetric}
\end{equation}
its energy and Bekenstein-Hawking entropy are
\begin{equation}
E_s=M,~~S_s=\frac{A}{4}=4\pi M^2, \label{EESchBH}
\end{equation}
where $A$ is the area of event horizon located at $r_h=2M$, and $c=G=\hbar=k_B=1$ have been assumed for the convenience here and in the following. Therefore, if there is a condensation of the Schwarzschild black hole among the radiation gas, the total energy and entropy inside the isolated box will be
\begin{equation}
E=M+aVT^4,~~S=4\pi M^2+\frac{4}{3}aVT^3, \label{EETotal}
\end{equation}
where the volume of black hole has been neglected.

Note that, for this isolated system, the total energy $E$ should be conserved, and hence the stable equilibrium state should have the maximal total entropy $S$. After setting $x=M/E$ and $y=(1/3\pi)(aV/E^5)^{1/4}$, one will find that maximizing the total entropy $S$ is equal to maximize the function
\begin{equation}
f(x)=x^2+y(1-x)^{3/4}, \label{Equation}
\end{equation}
where the range of independent variable $x$ is $[0,1]$.

For the convenience, we have drawn a simple diagram for this function $f(x)$ with respect to $x$ by choosing different constants $y$ in the Fig.~\ref{fig1}:
\begin{figure}[H]
\centering
\includegraphics[width=4in,height=3in]{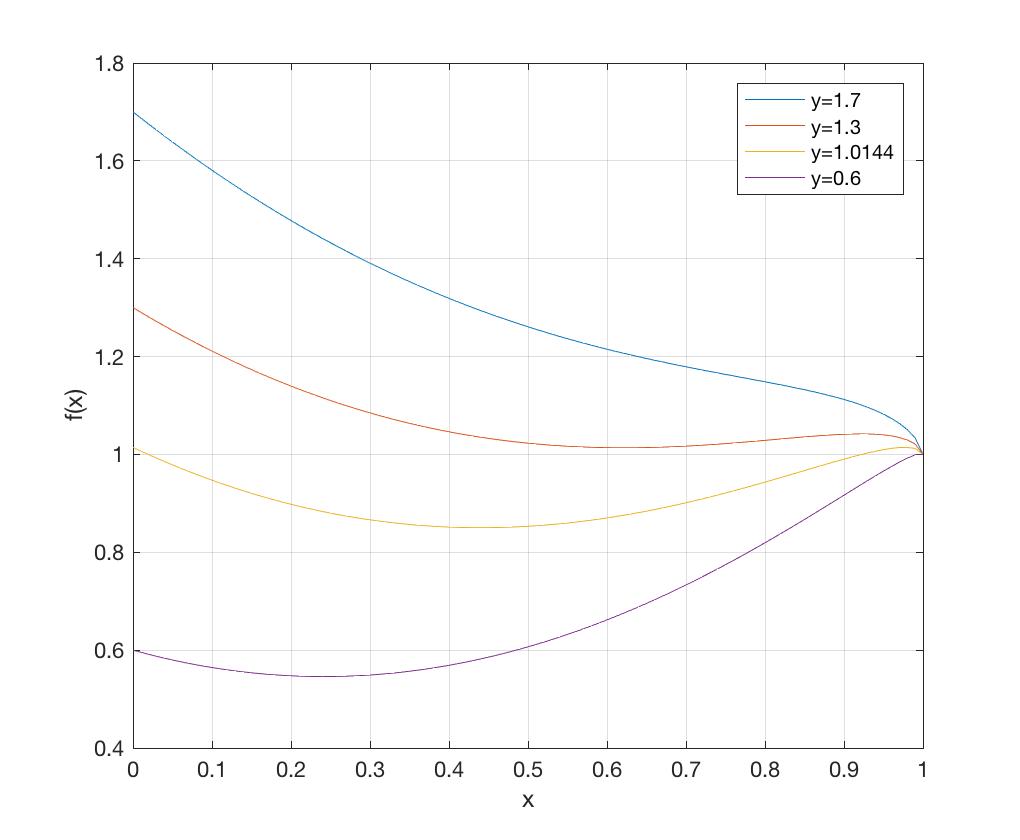}
\caption{The diagram of function $f(x)$ with respect to $x$ by choosing different constants $y$.}
\label{fig1}
\end{figure}

The main conclusions are:

(1) For $y>1.4266$, the function $f(x)$ is monotonously decreasing and there are no turning points, thus the maximum value
of $f(x)$ is at $x=0$, which means that the equilibrium configuration is just the pure radiation and there is no black hole.

(2) For $1.4266>y>y_c=1.0144$, there are two turning points, a local minimum for $x<4/5$ and a local maximum for $x>4/5$. However, the global maximum of $f(x)$ is still given by $x=0$, which means that the equilibrium configuration is still the pure radiation.

(3) For $y<y_c=1.0144$, there are also two turning points, but the global maximum of $f(x)$ given by $x> x_c =  0.97702$, which means that the stable equilibrium configuration consists of the Schwarzschild black hole and black-body radiation, and the temperature of radiation will be same as the temperature of Schwarzschild black hole $T=1/(8\pi M)$.

\begin{figure}[H]
\centering
\includegraphics[width=4in,height=3in]{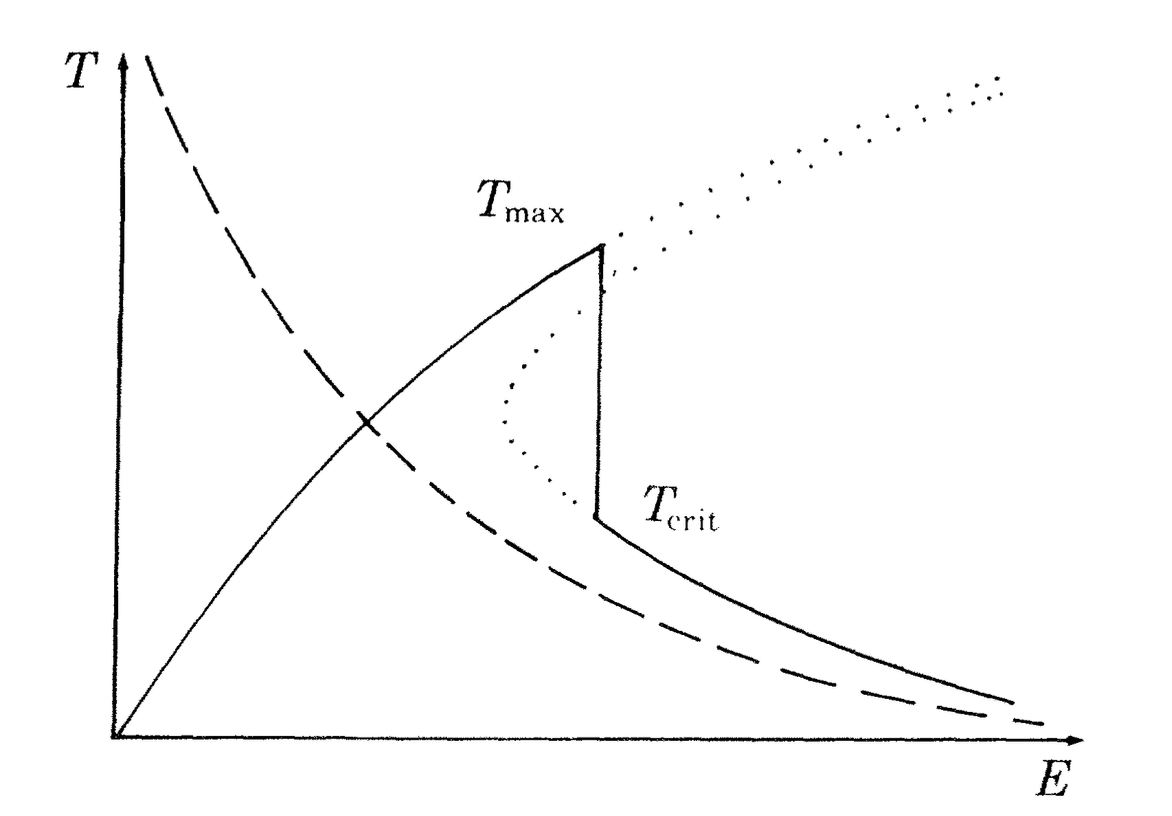}
\caption{The $T-E$ diagram in the Schwarzschild case.}
\label{fig2}
\end{figure}

In addition, if one fixes the volume $V$ of the isolated box and the Stefan's constant $a$, the $T-E$ diagram of this isolated system has been drawn by G.W Gibbons and M.J Perry in the Fig.~\ref{fig2}. In this figure, the solid line represents the stable equilibrium state, while the dotted lines represent unstable state and the dash line represent the pure black hole state.

From this figure, one can easily see that there is only the pure radiation if the total energy $E$ is low, since the constant $y=(1/3\pi)(aV/E^5)^{1/4}$ is very high, which is also consistent with the physics that the black hole will not be formed if the energy density is not high enough. However, when the total energy $E$ becomes bigger, the constant $y$ will be lower down to the critical value $y_c=1.0144$. If more energy $E$ is added, the black hole will be condensed among the radiation gas, i.e., the stable equilibrium configuration consists of the Schwarzschild black hole and black-body radiation. Therefore, there is an upper temperature bound $T_{max}$ for this isolated system with fixed volume $V$
\begin{equation}
T_{max}=\frac{1}{(3\pi y_caV)^{1/5}},
\end{equation}
while the upper temperature bound or critical temperature of black hole is
\begin{equation}
T_c=\frac{1}{8 \pi x_c}(3 \pi y_c)^{4/5}(aV)^{-1/5}.
\end{equation}
It should be emphasized that this temperature of black hole $T_c$ is usually not equal to the temperature of the rest of radiation gas during its condensation, i.e. usually lower than the temperature of radiation gas, and hence this configuration state is in fact not stable. Therefore, the black hole will become bigger, and finally the isolated system will reach the stable configuration state, i.e., the temperature of condensed black hole is equal to the temperature of radiation gas. Note that, during this non-equilibrium process, this isolated system with fixed volume $V$ will keep the same total energy $E_c=aVT_{max}^4$, and its temperature will be lower to the final value $T_{crit}$, i.e. the temperature of this isolated system in the stable configuration state and $T_{crit}<T_c<T_{max}$. After $E_c<E$, the temperature of this isolated in the stable configuration state will be lower than $T_{crit}$. Note that, an interesting prediction is that the radius of horizon is nonzero when the condensation of a Schwarzschild black hole occurs among the radiation gas.

\section{Effects from massive graviton on the equilibrium between the black hole and radiation gas in an isolated box}
In this section, we will investigate the effects from massive graviton on the equilibrium between the black hole and radiation gas in an isolated box. Note that, since the graviton is massive, the condensed black hole among the radiation gas is usually not the Schwarzschild black hole, which should be considered under the massive gravity theory. For the massive gravity theory, here we just use the dRGT massive gravity which has been proved to be ghost free. Therefore, we will first give a short introduction to the dRGT massive gravity and its Schwarzschild-like black hole solution, i.e. the spherically static black hole solution in the vacuum. Then, also using the conservation of energy and principle of maximal entropy for an isolated system, we will investigate the equilibrium between the Schwarzschild-like black hole and radiation gas in an isolated box. Since there are two parameters related to the graviton mass in the Schwarzschild-like black hole solution, here we just consider two simple cases related to the two parameters, which can be found to be analytically investigated. Moreover, we will find that these two simple cases can also express the interesting results. For example, the $T-E$ diagram is similar like the Schwarzschild case in the first case, while it is different in the second case.

\subsection{The dRGT Massive gravity and its Schwarzschild-like black hole solution}
The action of the dRGT massive gravity in an $(n+2)$-dimensional spacetime with the cosmological constant $\Lambda=-\frac{(n+1)n}{2\ell^2}$ is usually read as~\cite{Vegh:2013sk,deRham:2010ik,deRham:2010kj,Cai:2014znn}
\begin{equation}
\label{actionmassive}
S =\frac{1}{16\pi G}\int d^{n+2}x \sqrt{-g} \left[ R +\frac{n(n+1)}{L^2} +m^2 \sum^4_i c_i {\cal U}_i (g,\mathfrak{f})\right],
\end{equation}
where $\mathfrak{f}$ is a fixed symmetric tensor usually called the reference metric, $L$ is the radius of AdS$_{n+2}$ spacetime;
$c_i$ are constants,  $m$ stands for the graviton mass parameter, and ${\cal U}_i$ are symmetric polynomials of the eigenvalues of the $(n+2)\times (n+2)$ matrix ${\cal K}^{\mu}_{\ \nu} \equiv \sqrt {g^{\mu\alpha}\mathfrak{f}_{\alpha\nu}}$:
\begin{eqnarray}
\label{eq2}
&& {\cal U}_1= [{\cal K}], \nonumber \\
&& {\cal U}_2=  [{\cal K}]^2 -[{\cal K}^2], \nonumber \\
&& {\cal U}_3= [{\cal K}]^3 - 3[{\cal K}][{\cal K}^2]+ 2[{\cal K}^3], \nonumber \\
&& {\cal U}_4= [{\cal K}]^4- 6[{\cal K}^2][{\cal K}]^2 + 8[{\cal K}^3][{\cal K}]+3[{\cal K}^2]^2 -6[{\cal K}^4].
\end{eqnarray}
The square root in ${\cal K}$ means $(\sqrt{A})^{\mu}_{\ \nu}(\sqrt{A})^{\nu}_{\ \lambda}=A^{\mu}_{\ \lambda}$ and $[{\cal K}]=K^{\mu}_{\ \mu}=\sqrt {g^{\mu\alpha}\mathfrak{f}_{\alpha\mu}}$. After making variation of action with respect to the metric, the equations of motion (EoM) turns out to be
\begin{eqnarray}
R_{\mu\nu}-\frac{1}{2}Rg_{\mu\nu}-\frac{n(n+1)}{2L^2} g_{\mu\nu}+m^2 \chi_{\mu\nu}&=&8\pi G T_{\mu \nu },~~
\end{eqnarray}
where
\begin{eqnarray}
&& \chi_{\mu\nu}=-\frac{c_1}{2}({\cal U}_1g_{\mu\nu}-{\cal K}_{\mu\nu})-\frac{c_2}{2}({\cal U}_2g_{\mu\nu}-2{\cal U}_1{\cal K}_{\mu\nu}+2{\cal K}^2_{\mu\nu})
-\frac{c_3}{2}({\cal U}_3g_{\mu\nu}-3{\cal U}_2{\cal K}_{\mu\nu}\nonumber \\
&&~~~~~~~~~ +6{\cal U}_1{\cal K}^2_{\mu\nu}-6{\cal K}^3_{\mu\nu})
-\frac{c_4}{2}({\cal U}_4g_{\mu\nu}-4{\cal U}_3{\cal K}_{\mu\nu}+12{\cal U}_2{\cal K}^2_{\mu\nu}-24{\cal U}_1{\cal K}^3_{\mu\nu}+24{\cal K}^4_{\mu\nu}).~~
\end{eqnarray}
Since the background we are going to consider is $(3+1)$ dimension, and thus a general black hole solution in the vacuum has been found in~\cite{Cai:2014znn}
\begin{eqnarray}\label{metric}
ds^2&=&-f(r)dt^2+\frac{dr^2}{f(r)}+r^2h_{ij}dx^idx^j,\\
\label{fr} f(r)&=&k +\frac{r^2}{L^2}-\frac{m_0}{r}+\frac{c_1m^2r}{2}+c_2m^2,
\end{eqnarray}
where $h_{ij}dx^idx^j$ is the line element for the 2-dimensional spherical, flat or hyperbolic space with $k=1,~0$ or $-1$ respectively, and $m_0$ is related to the mass of the black hole. The reference metric now can have a special choice
\begin{equation}
\mathfrak{f}_{\mu\nu}=\text{diag}~\{0,0,h_{ij}\}.
\end{equation}
The Hawking temperature of this black hole solution can be easily found
\be
T_{BH}=\frac{\left(f(r)\right)'}{4\pi}\bigg|_{r=r_h}=\frac{1}{4\pi r_h}\left(k+\frac{3r_h^2}{L^2}+c_1m^2r_h+c_2m^2\right),
\ee
where $r_h$ is the event horizon of the black hole, i.e. the largest root of equation $f(r_h)=0$.

In our case, we are just interested in the Schwarzschild-like black hole solution with zero cosmological constant, therefore, the corresponding solution in (\ref{fr}) is
\begin{eqnarray}
\label{fr1} f(r)&=&1-\frac{2M}{r}+\frac{c_1m^2r}{2}+c_2m^2,
\end{eqnarray}
while its temperature, energy and entropy are
\begin{eqnarray}
T_{BH}&=&\frac{1}{4\pi r_h}(1+c_1m^2r_h+c_2m^2),\label{TBlackhole}\\
E&=&M,~~S=\pi r_h^2.
\end{eqnarray}
Note that, there are two parameters $c_1$ and $c_2$ related to the graviton mass in (\ref{fr1}), which can deduce the expression of entropy to be complicate. This complication will be further amplified during maximizing the total entropy of the isolated system, which can deduce the difficulty to analytically investigate the condition of phase transition, i.e. condensation of the Schwarzschild-like black hole among the radiation gas in an isolated box. Therefore, we will just consider two simple cases in the following, which have been found to be analytically investigated and also contained the interesting results.

\subsection{The first simple case: $c_1=0$}
For the case $c_1=0$, we can obtain the location of horizon
\begin{eqnarray}
r_h=\frac{2M}{c_2m^2+1}.
\end{eqnarray}
After the assumption of the condition $c_2m^2+1>0$, a positive radius of the black hole horizon can be obtained, and the total energy and entropy of the isolated system consisting of a Schwarzschild-like black hole and radiation gas can be easily calculated
\begin{eqnarray}
S&=&\pi r_h^2+\frac{4}{3}aVT^3=\frac{4\pi M^2}{(c_2m^2+1)^2}+\frac{4}{3}aVT^3, \\
E&=&M+aVT^4.
\end{eqnarray}
After introducing two parameters:~$x=M/E$,~$y=(aV/E^5)^{1/4}(c_2m^2+1)^2/3\pi$, we will find that maximizing entropy $S$ is also equal to maximize the function
\begin{eqnarray}
f(x)=x^2+y(1-x)^{3/4}.
\end{eqnarray}
Note that, this function is just same as the Schwarzschild case in (\ref{Equation}), and the only difference is that the parameter $y$ has been affected by the graviton mass. Therefore, the main conclusions are also same as the Schwarzschild case. But yet, just due to the same function and conclusions, an interesting result can be further found, i.e. effects from graviton mass on the equilibrium by comparison with the Schwarzschild case for the critical energy $E_c$.

\begin{figure}[H]
\centering
\includegraphics[width=4in,height=3in]{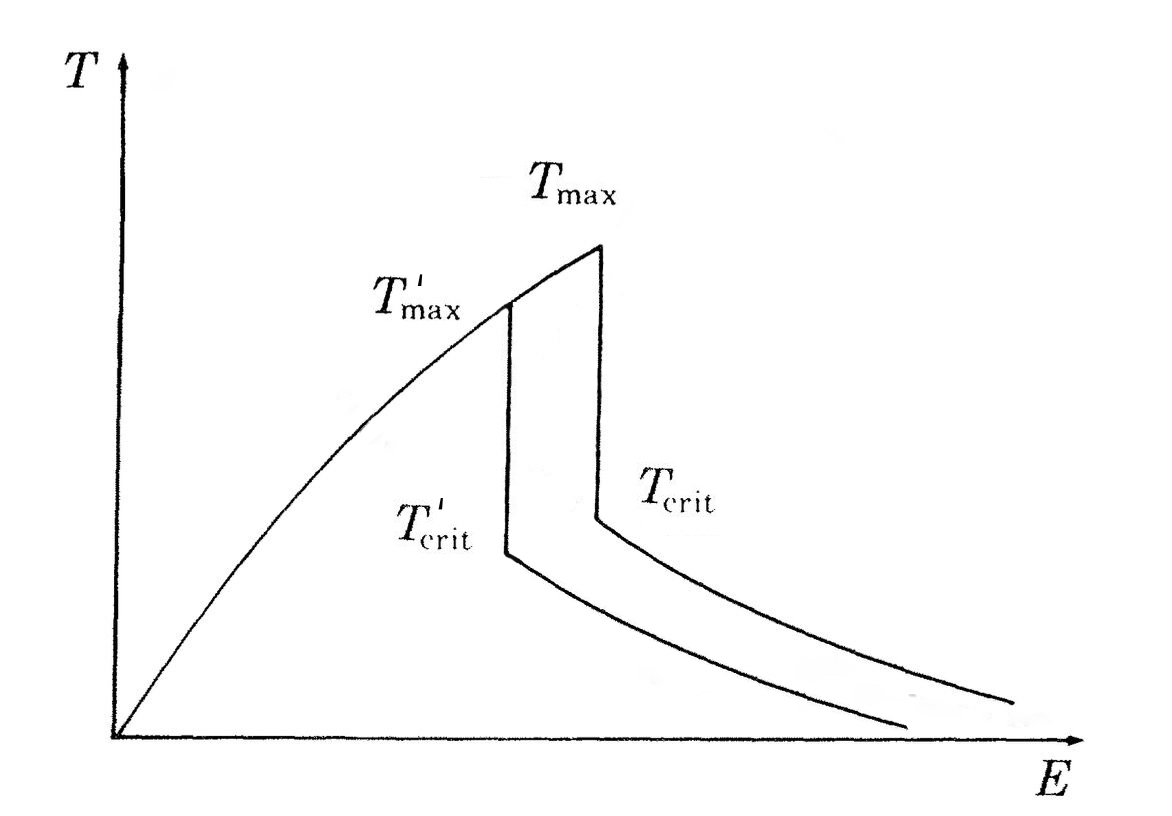}
\caption{The $T-E$ diagram in the dRGT massive gravity with $c_1=0$ and $c_2<0$. }
\label{fig3}
\end{figure}

For this case $c_1=0$ in the massive gravity, its corresponding critical energy is
\begin{eqnarray}
 E'_c=\left\{ \left[\frac{(c_2m^2+1)^2}{3\pi y_c}\right]^4 aV\right\}^{\frac{1}{5}},
\end{eqnarray}
while the critical energy in the Schwarzschild case is
\begin{eqnarray}
E_c=\left\{ \left[\frac{1}{3\pi y_c}\right]^4 aV\right\}^{\frac{1}{5}}.
\end{eqnarray}
Obviously, if $c_2<0$ and $0<c_2m^2+1<1$, we will obtain $E'_c<E_c$, which means that the graviton mass can increase the condensation of black hole with the same isolated box with fixed volume $V$. On the other hand, if $c_2>0$, then $c_2m^2+1>1$ and $E'_c>E_c$, which means that the graviton mass can suppress the the condensation of black hole. Note that, these interesting predictions can be also seen in the $T-E$ diagram under the fixed volume $V$ of the isolated box, where we have drawn the diagram in the case $c_1=0, c_2<0$ in the Fig. {\ref{fig3}}. In this figure, $T'_{max}$ and $T'_{crit}$ represent the two different turning points respectively compared with the Schwarzschild case, while the $T-E$ diagram in the other case $c_1=0, c_2>0$ has not been drawn since it is very similar with this diagram.

\subsection{The second simple case: $c_2m^2+1=0$}
For this case with $c_2m^2+1=0$, the radius of horizon $r_h$ is also simple
\begin{eqnarray}
r_h=\sqrt\frac{4M}{c_1m^2},
\end{eqnarray}
where $c_1>0$ has been assumed to obtain a positive radius of horizon. Then, the total energy and entropy of the isolated system can be calculated
\begin{eqnarray}
S&=&\pi r_h^2+\frac{4}{3}aVT^3=\frac{4\pi M}{c_1m^2}+\frac{4}{3}aVT^3, \label{EQEcase1}\\
E&=&M+aVT^4.\label{EQEcase2}
\end{eqnarray}
From Eq.(\ref{EQEcase2}), we can obtain $T=\left( \frac{E-M}{aV} \right)^{1/4}$ which is substituted into Eq.(\ref{EQEcase1}). Hence, we will have
\begin{eqnarray}
S=\frac{4\pi M}{c_1m^2}+\frac{4}{3}aVT^3 =\frac{4\pi E}{c_1m^2}\left[ \frac{M}{E}+\frac{1}{3\pi}\left( \frac{aV}{E} \right)^{1/4}c_1m^2\left(1-\frac{M}{E}\right)^{3/4}\right].
\end{eqnarray}
Similarly, after introducing two parameters: $x=M/E$, $y=(aV/E)^{1/4}c_1m^2/(3\pi)$, then we will find that maximizing entropy $S$ is equal to maximize the function
\begin{eqnarray}
f(x)=x+y(1-x)^{3/4}. \label{Fcase2}
\end{eqnarray}
For this function $f(x)$ in (\ref{Fcase2}), the main conclusions are:

(1) For $y>\frac{4}{3}$, the function $f(x)$ is monotonously decreasing, thus the maximum value
of $f(x)$ is at $x=0$, which means that the stable equilibrium configuration is just the pure radiation and there is no black hole.

(2) For $0<y<\frac{4}{3}$, there is just one turning point, a maximum at specific $x_c=1-\left( \frac{3}{4}y \right)^4$, which means that the stable equilibrium configuration consists of the Schwarzschild-like black hole and black-body radiation.

If the volume $V$ of the isolated box is fixed, the above conclusions can also have the corresponding physical meaning in the following. Suppose that we add more energy $E$ into the isolated box with a fixed volume $V$ from $E=0$, the stable equilibrium configuration is just the pure radiation until the parameter $y=(aV/E)^{1/4}c_1m^2/3\pi$ is lower to the critical value $y_c=\frac{4}{3}$. After adding more energy, the $y$ will become smaller than $y_c=\frac{4}{3}$, and then the black hole will be condensed among the radiation gas, i.e., the stable equilibrium configuration consists of the Schwarzschild-like black hole and black-body radiation. Therefore, there is also a critical energy
\begin{eqnarray}
E_c=aV\left( \frac{c_1m^2}{4\pi} \right)^4,
\end{eqnarray}
and the corresponding temperature for this isolated system is
\begin{eqnarray}
T_{max}=\frac{c_1m^2}{4\pi}.
\end{eqnarray}

Note that, in this case with $c_2m^2+1=0$, a very interesting and impressive result is that the above temperature is just same as the temperature of Schwarzschild-like black hole condensed in $(\ref{TBlackhole})$:
\begin{eqnarray}
T_h=\frac{1}{4\pi r_h}(1+c_1m^2r_h+c_2m^2)=\frac{c_1m^2}{4\pi},
\end{eqnarray}
which means that the temperature of this isolated system will keep same after more energy $E$ is added into, and there is no non-equilibrium process after the condensation of the black hole in this case. Therefore, after the total energy $E_c<E$, the stable equilibrium configuration consists of the Schwarzschild-like black hole and black-body radiation, and the main change is just that the maximum point $x_c=1-\left(\frac{3}{4}y\right)^4$ will become bigger and closer to 1 from 0, i.e., the percentage of black hole's energy in the total energy will become bigger. A most interesting and impressive prediction is that the radius of horizon will become bigger from zero when the Schwarzschild-like black hole is condensed among the radiation gas, which is very different from the Schwarzschild case or the above case. This difference can be also clearly seen in the corresponding new $T-E$ diagram with fixed volume $V$ in this case, which has been drawn in the Fig.~\ref{fig4}:
\begin{figure}[H]
\centering
\includegraphics[width=4in,height=3in]{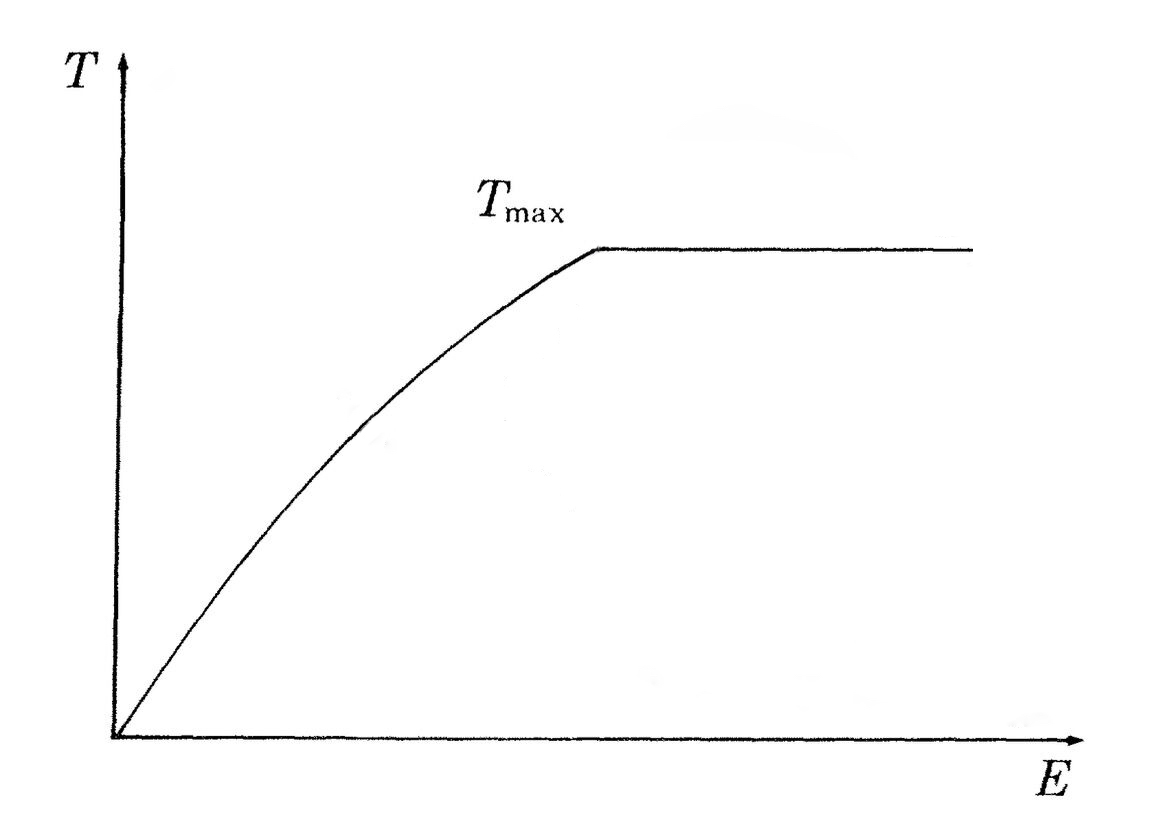}
\caption{The $T-E$ diagram in the dRGT massive gravity with $c_2m^2+1=0$ and $c_1>0$. }
\label{fig4}
\end{figure}

\section{Conclusion and discussion}
In this paper, we mainly investigate the effects from massive graviton on the equilibrium between the black hole and radiation gas in an isolated box. Since the graviton is massive, the underlying gravity theory should be massive gravity, and hence the condensed black hole among the radiation gas will usually not be the Schwarzschild black hole. Therefore, we have taken the dRGT massive gravity into account, i.e. a ghost free massive gravity, while the Schwarzschild-like black hole solution in this dRGT massive gravity has already been found and investigated much. After using the conservation of total energy and maximizing the total entropy for this isolated system, we find that two parameters $c_1$ and $ c_2$ are crucial to investigate the effects from massive graviton on the equilibrium analytically. For the simplicity, we just consider two simple cases which can be found to have analytically investigations on the conditions of phase transition or the $T-E$ phase diagram. Moreover, some interesting and new results have been obtained, too. For example, in the first case with $c_1=0$, although the $T-E$ phase diagram is similar as the Schwarzschild case, however, we can further conclude that the graviton mass can suppress or increase the condensation of black hole which is dependent on the value of $c_2$. For the other case with $c_2m^2+1=0$, a new $T-E$ phase diagram turns out. Moreover, an interesting and important prediction is that the condensation of black hole just increases from the zero radius of horizon in this case, which is very different from the Schwarzschild black hole case.

A direct interesting question is that whether there are other analytical investigations for different choices of $c_1$ and $c_2$. Moreover, since $c_1$ and $c_2$ are just proposed as two constants in the dRGT massive gravity, there is no further information for these two parameters in this gravity theory. Therefore, maybe some experiments or astronomical phenomenons can be related to this equilibrium between the Schwarzschild-like black hole and radiation gas in an isolated box, which can be used to test the dRGT massive gravity and further extract the information of these two parameters. However, since the Birkhoff's theorem may be broken in the dRGT massive gravity, more cases need be considered, i.e. other static black hole condensed among the radiation gas~\cite{Volkov:2013roa,Tasinato:2013rza,Babichev:2015xha}. In addition, the information of the reference metric need be also carefully investigated, since different reference metric may also deduce the different static black hole in massive gravity~\cite{Volkov:2013roa,Tasinato:2013rza,Babichev:2015xha}. On the other hand, the cosmological constant can be also taken into account during the investigations on the equilibrium. For the positive cosmological constant, since our universe may be the de Sitter spacetime during its inflation time or late time with acceleration, and hence the cosmological horizon can be naturally considered as the wall of the isolated box. For the negative cosmological constant, the static black hole solutions are usually asymptotical to the Anti-de Sitter spacetime, which have been found to have many interesting results, such as the well-known Hawking-Page transition~\cite{Hawking:1982dh}. Moreover, according to the AdS/CFT correspondence, the static black hole solutions asymptotical to the Anti-de Sitter spacetime can have many interesting results on the dual field theory, such as the holographic superconductor in the holographic duality in the condensed matter physics~\cite{Zaanen:2015oix}. In addition, the isolated box may be considered as the cutoff surface related to the energy scale of renormalization group (RG) flow for the strongly coupling holographic fluid~\cite{Bredberg:2010ky,Cai:2011xv,Niu:2011gu,Bai:2012ci,Hu:2013lua}. Therefore, during the investigations on the equilibrium between the black hole and radiation gas in an isolated box, it will be also an interesting issue to be further studied by taking the cosmological constant into account.

\section*{Acknowledgments}
Y.P Hu thanks a lot for Profs. Hong-Sheng Zhang and Hai-Qing Zhang's valuable comments. This work is supported by the Fundamental Research Funds for the Central Universities under grant No. NS2015073.

\end{document}